\documentclass[12pt]{article}

\usepackage{amsthm,amsmath,natbib}
\usepackage{amsfonts}
\usepackage{graphicx,bm}
\usepackage{lscape}
\usepackage{rotating}
\usepackage{setspace}
\usepackage[most]{tcolorbox}
\usepackage[T1]{fontenc}
\usepackage{tgpagella}
\usepackage{quoting,xparse}
\NewDocumentCommand{\bywhom}{m}{% the Bourbaki trick
  {\nobreak\hfill\penalty50\hskip1em\null\nobreak
   \hfill\mbox{\normalfont(#1)}%
   \parfillskip=0pt \finalhyphendemerits=0 \par}%
}
\NewDocumentEnvironment{pquotation}{m}
  {\begin{quoting}[
     indentfirst=true,
     leftmargin=\parindent,
     rightmargin=\parindent]}
  {\bywhom{#1}\end{quoting}}

\begin{document}

\begin{titlepage}% Definition of title page:
\title{The Early Roots of Statistical Learning in the Psychometric Literature: A review and two new results}
\author{Mark de Rooij, Bunga Citra Pratiwi, Marjolein Fokkema, \\Elise Dusseldorp, Henk Kelderman\\ 
Leiden University\\ 
Institute of Psychology\\
Methodology \& Statistics Unit\\
PO Box 9555\\ 
2300 RB Leiden\\
The Netherlands\\
rooijm@fsw.leidenuniv.nl}
\date{}
\maketitle

\begin{abstract}
Machine and Statistical learning techniques become more and more important for the analysis of psychological data. Four core concepts of machine learning are the bias variance trade-off,  cross-validation, regularization, and basis expansion. We present some early psychometric papers, from almost a century ago, that dealt with cross-validation and regularization. From this review it is safe to conclude that the origins of these lie partly in the field of psychometrics. From our historical review, two new ideas arose which we investigated further: The first is about the relationship between reliability and predictive validity; the second is whether optimal regression weights should be estimated by regularizing their values towards equality or shrinking their values towards zero. In a simulation study we show that the reliability of a test score does not influence the predictive validity as much as is usually written in psychometric textbooks. Using an empirical example we show that regularization towards equal regression coefficients is beneficial in terms of prediction error.  
\end{abstract}
\end{titlepage}

\section{Introduction}

For the analysis of psychological data more and more often statistical learning (a.k.a. machine learning or predictive modelling- we use the terms interchangeably) techniques are advertised, see for example \cite{YarkoniWestfall2017, Tonidandelea2016, Putkaea2018, Chapmanetal2016, McNeish2015}. We, the authors, think that this pronounced emphasis on these techniques is in general a good idea. While studying the concepts we found that many ideas go back to the early psychometric literature. 

Four core elements in much of the statistical learning literature are \emph{bias variance trade-off}, \emph{cross-validation}, \emph{regularization}, and \emph{basis expansion}. These four elements are important to find an optimal model for a given data set.  We provide a brief overview below, more detailed discussions are provided in \cite{HTF2009} or \cite{Berk2008}.

The bias variance trade-off deals with avoiding over- and underfitting. With overfitting we capture too much noise, with underfitting too little signal. In both cases, the fitted model does not generalize well to new observations from the same population. With an overfitted model the bias reduces to zero, that is, would we fit this model to repeated samples from the population the average of all models is representative for the true population model. However, this model may have large variance, meaning that the estimated parameters from sample to sample vary a lot. On the other hand with an underfitted model the variance from sample to sample decreases at the cost of an increased bias. 

Machine learners try to find the optimal model by avoiding underfitting and overfitting. In practice this often boils down to fitting series of models and selecting the optimal one. The selection of an optimal model is done using \emph{cross-validation}. This technique trades-off bias and variance by finding the model that best generalizes to unseen data.

Early researchers focussed largely on linear models because they were computationally feasible in the pre-computer era. Thinking about over- and underfitting, the linear model might be too complex capturing noise or the linear model might be too simple capturing little signal. In the latter case, the relationship in the population is nonlinear and the linear model does not capture this. In the process of finding an optimal model and starting with a linear model, two directions can be distinguished: more and less flexible models. 

Let us start with less flexible models. Modern statistical learning techniques add a penalty part to the (least squares) loss function that is minimized in order to obtain regression weights. In lasso regression the penalty part consists of the sum of absolute values of the regression weights \citep{Tibshirani1996}, in ridge regression the penalty part consists of the sum of squared values of the regression weights \citep{HoerlKennard1970}, and in elastic net it is a combination of the lasso and ridge penalty \citep{ZouHastie2005}. Another way of understanding these \emph{regularization} techniques is stating that the method searches for least squares estimates but only in a small region of the parameter space; the form of the region is defined by the penalty function; see \cite[][page 22, 58]{Sparsity2015} for graphical examples of these regions.

We can also make the model more flexible. One obvious possibility is by adding polynomial terms of the predictor into the regression equation, so not only using $X$, but also $X^2$, $X^3$, etc. Such an operation is called \emph{basis expansion}, because a single predictor is blown up to multiple predictors. Other basis expansion operations are spline regression models or kernel regression models.

In the next section we delve in the history of psychometrics with a focus on cross-validation, the bias variance trade-off, and regularization. We will cite and quote from several papers starting in the 1920s and we bounded (regularized) our search till 1980. We will see that a lot of ideas originated from this early psychometric literature. Moreover, investigating these old papers and relating conclusion from that work to current psychometric knowledge brought us with two new questions/ideas which we will discuss in Section 3. We conclude this paper with some discussion and ideas for teaching and future work. 

\section{History}\label{sec:hist}

In this section we present some early papers about cross-validation and regularization in the psychometric literature. Most of this literature is concerned with prediction; more specifically the question how to combine item scores of a psychological test or test scores from a test battery into a composite score to select persons for a job, or students for a school. We provide quotes from these papers to give some idea of the research tendencies at that time. We tried to find origins in the psychometric literature, but are by no means sure we found them. Moreover, we probably missed some contributions for which we apologise in advance. This section is divided in two subsections, one about the early roots of cross-validation and the other about the early roots of regularization. 

\subsection{Cross-validation}

\begin{pquotation}{Selmer C. Larson, 1931}
`It has been recognized by theoretical statisticians for some time that when a coefficient of multiple correlation ($R$) is derived for a given set of data, its value is likeliy to be deceptively large. If the computations have been correct, the value will hold rigidly for the set of data from which the regression equation was derived. If, however, the equation should be applied to a second set of data, even though strictly comparable, it has been supposed that the yield in this latter case would, except for errors due to sampling, be less than in the first. Moreover, it has been supposed that the more variables contained in the regression equation, the greater this shrinkage will be. This is particularly significant because ordinarily the practical employment of a regression equation involves its use with data other than those from which it was derived. ' 
\end{pquotation}

Because of the lack of automated computational power, \cite{Larson1931} works out an adjustment to the multiple correlation coefficient. This has been the start of a long series of papers with adjustments to the $R$ \citep{Wherry1951, Darlington68, Browne1975a, Browne1975b, Rozeboom1978, Claudy1978}. To derive these adjustments distributional assumptions need to be made, which is a severe drawback of these indices, see \nocite{Herzberg1969}

\begin{pquotation}{Paul A. Herzberg, 1969}
`However, these formulas require assumptions which are often difficult to satisfy and, therefore, many early investigators estimated the population correlation by applying to a second sample the regression weights calculated in an original sample. They found that the correlation between the regression function and the criterion in the second sample was less than the original sample multiple correlation. This technique became known as cross-validation of the predictor weights or simply as cross-validation (Mosier, 1951). The correlation in the second sample is called the cross-validity. '
\end{pquotation}

In the 1950 psychometric society meeting there was a symposium about cross-validation with Mosier, Cureton, Katzell, and Wherry as presenters. Their contributions appeared in the \emph{Educational and Psychological Measurement} of 1951. An important contribution was made my Mosier, who basically was the first who introduced $K$-fold cross-validation, with $K = 2$ to overcome the main problem with the loss off efficiency in split sample validation. Two quotes from these papers read:

\begin{pquotation}{Charles I. Mosier, 1951}
`If the combining weights of a set of
predictors have been determined from the statistics of one
sample, the effectiveness of the predictor-composite \emph{must} be
determined on a separate, independent sample.'
\end{pquotation}

\begin{pquotation}{Edward E. Cureton, 1951}
`Psychologists have long been accustomed to grinding out multiple-regression equations and asserting that the regression
coefficients so obtained are the best predictor weights that can be determined from the sample data. In one sense they are
correct. The least-squares regression weights are the best weights to use in predicting the criterion scores of the validation-
sample subjects. Their use maximizes the multiple correlation \emph{in that particular sample}. It does so by giving optimal
weights to \emph{everything} which, in that sample, will contribute to prediction. ``Everything'' includes, however, the sampling errors
in the validation-sample data. Hence the least-squares process \emph{over-fits}; it fits the errors as well as the systematic
trends in the data.'
\end{pquotation}

We see that the origins of cross-validation lie far back in history and as a matter of fact in the psychometric literature. We also note that the term \emph{overfitting} was used already in 1951. These papers indicate that overfitting is more pronounced for small samples, many predictor variables, and small effect sizes. These conditions were then and now the rule in psychological research. 

\subsection{Regularization}

Regularization is commonly understood as making linear models less flexible by imposing constraints. Popular machine learning methods are lasso regression \citep{Tibshirani1996} and elastic-net regression \citep{ZouHastie2005}. By imposing a penalty in the fitting procedure the variance from sample to sample is decreased and hopefully this trades off against the increased bias. In machine learning, regression weights are shrunken towards zero. In the psychometric literature we found the following quote about choosing the regression weights and verifying the prediction accuracy: \nocite{Cureton1951}

\begin{pquotation}{Edward E. Cureton, 1951}
`Hence, we will probably achieve a higher aggregate correlation in a second sample if we weight the standard predictor scores equally,
rather than by their beta regression coefficients as determined from the validation sample.'
\end{pquotation}

In a similar fashion, \cite{LawsheSchucker1959} investigated the following four weighting schemes  
\begin{itemize}
\item simple addition of raw scores
\item weighting raw scores by their standard deviation
\item weighting of raw scores by the inverse of the standard deviation
\item weighting by the least squares regression weights
\end{itemize}
and found NO evidence in favor of one of them over the others. Similar results can be found in \cite{Schmidt1971} and \cite{Wainer76}. The idea of equal weighting versus least squares weighting goes back to the early beginnings of psychometrics, see for example \cite{Burt1950} who quoted 

\begin{pquotation}{Frank N. Freeman, 1926}
`weighting has come to be far less commonly used than it was a number of years ago.'
\end{pquotation}
and 
\begin{pquotation}{Joy P. Guilford, 1936}
`weighting is not worth the trouble.'
\end{pquotation}
\nocite{Freeman1926, Guilford1936}.

The four weighting schemes described above can be understood using the bias variance trade-off: Equal weighting obviously does not give an unbiased estimate of the true regression equation, but because the data are not used for estimation the sample to sample variance is zero. This may be beneficial in some data analysis situations while harmful in others. This caused an argument between \cite{Wainer76} and \cite{Pruzek78}: Wainer claimed equal regression weights are beneficial in almost any circumstance, while Pruzek and Frederick claimed that only in very limited situations equal weighting is beneficial.   

\cite{Darlington78} was the first author who investigated these attempts in terms of the bias variance trade-off. He entitled his paper ``reduced variance regression'' in order to pay attention to the positive aspect (reducing variance) instead of the negative (increasing bias). For a simple regression model with population parameter $\beta$ and $b$ the least squares unbiased estimator of $\beta$ he showed that the expected squared error of a shrunken estimator $sb$ wih $0 \leq s \leq 1$ equals
\begin{eqnarray*}
\mathbb{E} (sb - \beta)^2 = s^2\sigma_b^2 + (1-s)^2\beta^2,
\end{eqnarray*}
where the first term represents the variance (which is equal to zero when $s=0$) and the second term the squared bias (which is equal to zero when $s = 1$). Minimizing this function for $s$ gives $s = \frac{1}{1 + \sigma^2_b/\beta^2}$ which is always smaller than 1, i.e. the least squares estimator is never optimal. Darlington further worked out this latter formula to 
$$
s = \frac{1}{1+\frac{(1-r^2)/r^2}{n-3}},
$$
where $r$ represents the population correlation between the predictor and response, which clearly shows the dependency of the amount of shrinkage on the strength of the population relationship and the sample size. The parameters on which $s$ depends are generally unknown and therefore the adjustment cannot readily be made in data analysis. 

After this analysis of simple regression, Darlington investigated results for multiple regression and discussed the Stein estimator and ridge regression in terms of bias and variance. Moreover, he gave an interpretation of ridge regression in terms of principal component regression and validity concentration. 

\section{Two New Ideas}

While studying the statistical learning papers and the early psychometric papers about cross-validation and regularization, two questions arose. The first question is about the relationship between reliability of the test score and its predictive validity. The second is about regularization towards equal regression weights, as opposed to shrinkage towards zero as in the lasso regression model. 

\subsection{Idea I: Validity - Reliability relationship}

In standard psychometric textbooks we often read that the reliability is an upper bound to the criterion- or predictive validity and that with decreasing reliability also the predictive validity goes down. From the recent machine learning literature and from the papers cited in section \ref{sec:hist} we conclude that to objectively say something about predictive validity we should not use the multiple $R$ or another in-sample measure, but a cross-validated measure. Furthermore, in order to optimize predictive accuracy regression parameters should be shrunken. The question now arises how reliability, shrinkage, and cross-validated prediction accuracy (or error) are related. 

Suppose we have a test score $X$ which is not a perfectly reliable score, i.e. $X = T + E$ where $T$ and $E$ are uncorrelated. The reliability of the test score can then be defined as
$$
\rho = \sigma^2_T/\sigma^2_X = \sigma^2_T/(\sigma^2_T + \sigma^2_E), 
$$
where $\sigma^2$ denotes the variance. 

Furthermore, suppose the regression on the observed test score is
\begin{eqnarray*}
Y &=& \alpha + \beta X + \epsilon \\
 &=& \alpha + \beta (T + E) + \epsilon \\
 &=& \alpha + \beta T  + \beta E + \epsilon \\
 &=& \alpha + \beta T  + \epsilon^*,
\end{eqnarray*}
where $\epsilon^* = \beta E + \epsilon$; the error variance increases due to unreliability of the test score. In terms of reduced variance regression \citep{Darlington78}, we note that the optimal shrinkage factor for unreliable test scores $s$ becomes smaller, i.e. more shrinkage is needed. One may, however, wonder what the effect is of reliability on predictive validity once this shrinkage is taken into account. To answer this question we set up a simulation experiment. 

We generate a calibration set and a validation set, also known as training and test set, respectively. In this experiment the true score ($T$) is distributed following a standard normal distribution. The criterion ($Y$) is related to the true score by 
$$
Y = \alpha + \beta T + \epsilon,
$$
with $\alpha = 0$ and $\epsilon$ is drawn from a normal distribution with standard deviation equal to $\sqrt{1-\beta^2}$. The correlation between the true scores and criterion scores in the population equals $r  \in\{0.20,0.25,0.30,0.35,0.40\}$, which we call the effect size. Observed test scores are obtained by
$$
X = T + E,
$$
where $E$ is a normally distributed variable with mean zero and variance $\sigma^2_E$. In our case, with $\sigma_T^2 = 1$, this variance ($\sigma^2_E$) is related to the reliability ($\rho$) through 
$$
\sigma^2_E = (1 - \rho)/\rho.
$$
We investigated reliabilities ranging from 1 to 0.5, i.e., from a perfectly reliable test till an unreliable test. Calibration data were generated with varying samples sizes of 25, 50, 100, and 200, respectively. Validation data were generated following the same model, but with a sample size of 1000. We replicated the procedure 1000 times. 

In the calibration set we use 10 fold cross-validation to find an optimal value of the shrinkage parameter $s$, that is, the value that minimizes the prediction error. Then we fitted the linear regression model with this value of the shrinkage parameter to the complete calibration set in order to find an estimated intercept and slope. Using the estimated intercept and slope from the calibration set and the observed test scores in the validation set, we derive predicted criterion scores in the validation set. The prediction error is defined as the mean squared difference between the predicted and the observed criterion scores in the validation set. 

\begin{sidewaysfigure}
\begin{center}
\includegraphics[width = 0.9\textwidth]{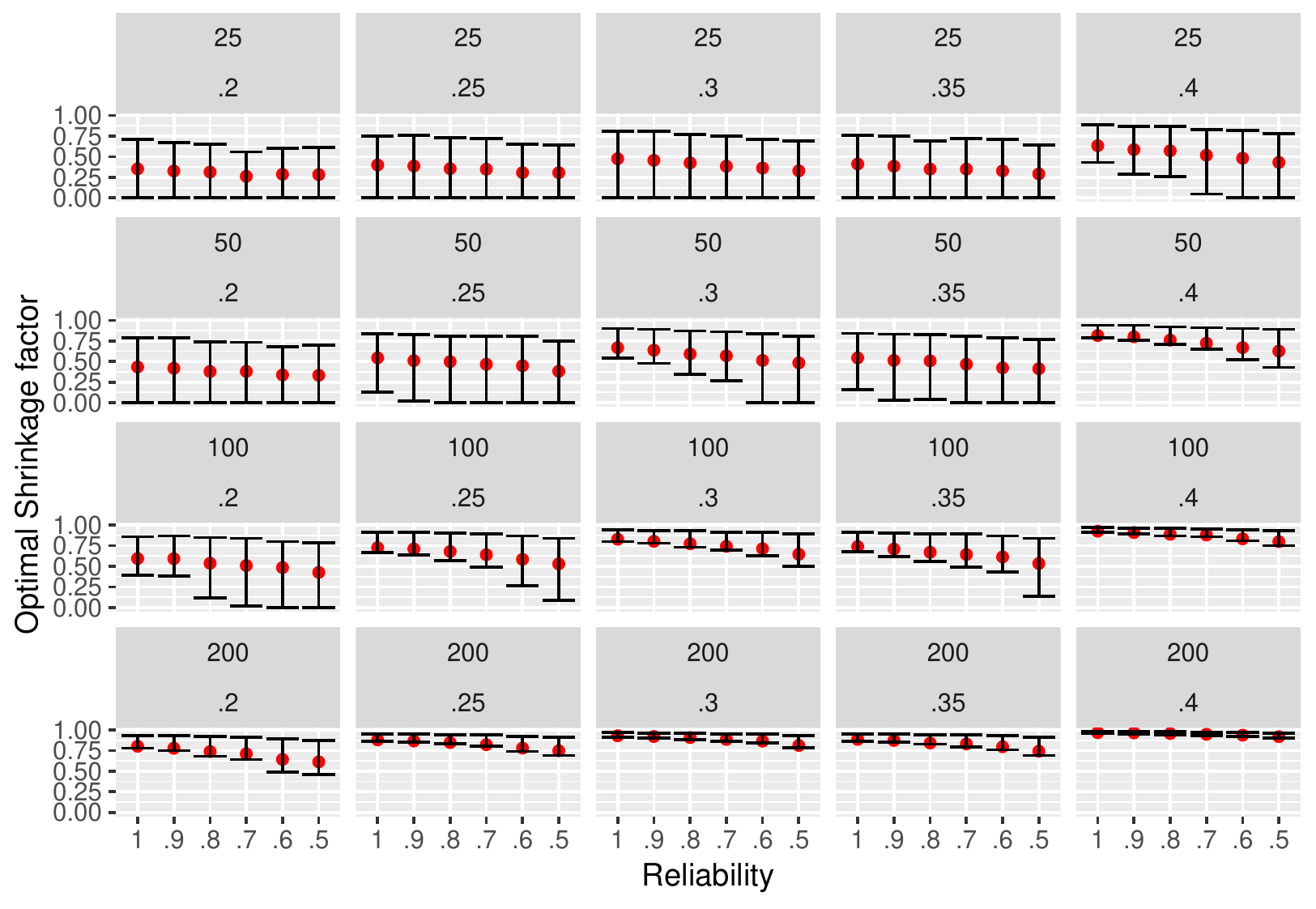}
\end{center}
\caption{Optimal Shrinkage factor ($0 \leq s \leq 1$) against reliability for different sample sizes ($n \in \{25, 50,100,200\}$) and effect sizes ($r \in \{.20,.25,.30,.35,.40\}$).The error bars show the 25th and 75th quantile.}
\label{fig:fig3a}
\end{sidewaysfigure}

For every replication the optimal value for the shrinkage parameter and the corresponding prediction error is computed. 
In Figure \ref{fig:fig3a} we see that, as expected, with decreasing reliability the optimal shrinkage factor $s$ becomes smaller, i.e. more shrinkage is necessary. More specifically, we see that with small effect sizes and small sample sizes the optimal value for $s$ is very low and sometimes even 0 (in which case predictions are based only on the estimated intercept). With larger effect sizes or sample sizes the amount of shrinkage is lower (values for $s$ are closer to 1). 

\begin{sidewaysfigure}
\begin{center}
\includegraphics[width = 0.9\textwidth]{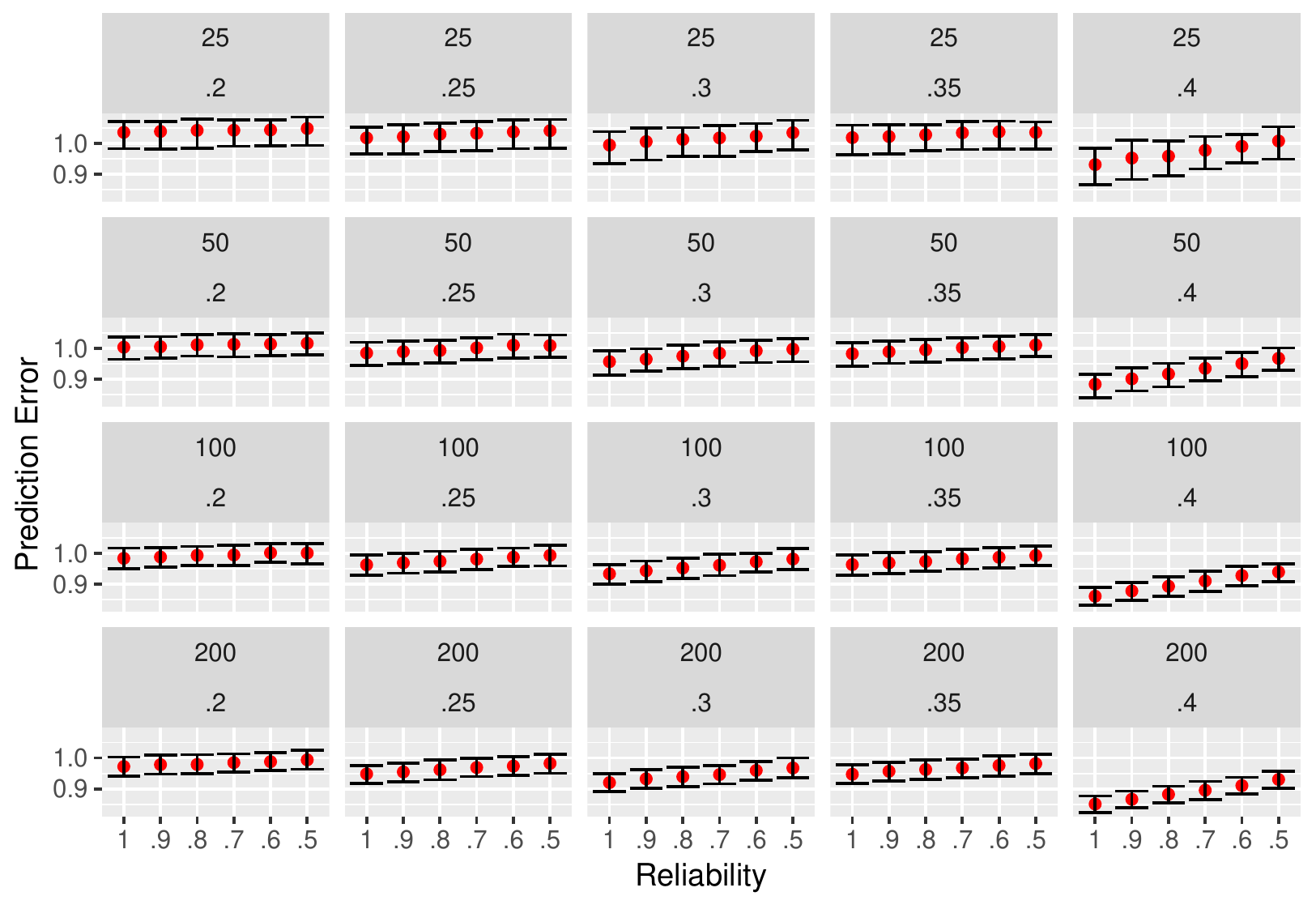}
\end{center}
\caption{Prediction error obtained with the optimal shrinkage factor against reliability for different sample sizes ($n \in \{25, 50,100,200\}$) and effect sizes ($r \in \{.20,.25,.30,.35,.40\}$). The error bars show the 25th and 75th quantile.}
\label{fig:fig3b}
\end{sidewaysfigure}

Figure \ref{fig:fig3b} shows the relationship between reliability and predictive validity. For small effect sizes ($r = 0.20, 0.25$), as they often occur in psychology, the effect of reliability on prediction error is minimal, no matter what sample size the prediction error is about equal. For larger effect sizes ($r = .40$) prediction error increases when reliability decreases (right hand sided frames). The latter effect is stronger for larger sample sizes.  

\subsection{Idea II: Shrinkage to equal regression weights}

In statistical learning the regression weights are often shrunken towards zero. In the early psychometric papers, in contrast, the weights were shrunken towards their mean. The idea arose that we can combine the two, i.e. use standard software for penalized regression models in order to shrink towards equal regression weights.

Therefore, let us define the regression model with equal weights as
\begin{eqnarray*}
Y &=& \alpha + \beta X_1 + \ldots + \beta X_p + \epsilon \\
&=& \alpha + \beta (X_1 + \ldots + X_p) + \epsilon \\
&=& \alpha + \beta X_+  + \epsilon, 
\end{eqnarray*}
where $X_+$ is the sum score for a subject. 

Also define the multiple regression model as
\begin{eqnarray}
Y &=& \alpha + \beta_1 X_1 + \ldots + \beta_p X_p + \epsilon. 
\label{eq:regression}
\end{eqnarray}
Here we have $p$ regression weights from which we can define $\beta = \frac{1}{p}\sum_j \beta_j$ (the average weight) and $\gamma_j = \beta_j - \beta$. Therefore the same multiple regression model can be written as
\begin{eqnarray*}
Y &=& \alpha + (\beta+ \gamma_1) X_1 + \ldots + (\beta+\gamma_p) X_p + \epsilon \\
&=& \alpha + \beta (X_1 + \ldots + X_p) + \gamma_1 X_1 + \ldots + \gamma_p X_p + \epsilon \\  
&=& \alpha + \beta X_+ + \gamma_1 X_1 + \ldots + \gamma_p X_p + \epsilon.  
\end{eqnarray*}
Furthermore, define $X_j = X_{\bullet} + Z_j$ with $X_{\bullet} = \frac{1}{p}X_{+}$ the person average score and replace in the equation above $X_j$ with $X_{\bullet} + Z_j$ to obtain
\begin{eqnarray}
Y &=& \alpha + \beta X_+ + \gamma_1 (X_{\bullet}+ Z_1) + \ldots + \gamma_p (X_{\bullet}+ Z_p) + \epsilon \nonumber \\
 &=& \alpha + \beta X_+ + (\gamma_1 +\ldots + \gamma_p) X_{\bullet}+ \gamma_1 Z_1 + \ldots + \gamma_p Z_p + \epsilon \nonumber \\
 &=& \alpha + \beta X_+ + \frac{1}{p}\gamma_+ X_+ + \gamma_1 Z_1 + \ldots + \gamma_p Z_p + \epsilon \nonumber \\
 &=& \alpha + \xi X_+  + \gamma_1 Z_1 + \ldots + \gamma_p Z_p + \epsilon.
\label{eq:constlasso}
\end{eqnarray}
Using these equations we rewrote the multiple regression model (equation \ref{eq:regression}) into another form where the predictors are a sum score ($X_+$) and the deviances of item scores ($Z_j$) towards the person mean item score (equation \ref{eq:constlasso}). These two models are equivalent in OLS terms, that is, they have the same amount of variance explained. This latter regression cannot be estimated using standard least squares procedures, because the predictor variables $Z_j$ are perfectly collinear (so we need to remove one of the $Z_j$'s). However, penalized regression models have no difficulty with multicollinearity; in fact, that is why they were designed in the first place. Therefore, we can apply $L_1$ or $L_2$ penalties on the $\gamma$-coefficients, and this can be done simply in standard software such as the glmnet package in R \citep{glmnet}. 

Let us verify how this penalizing towards an equal regression coefficient works in an empirical application. \cite{GarnefskiKraaij2007} describe a questionnaire for the assessment of cognitive emotion regulation. It consists of  36 five point Likert items measuring nine conceptually different subscales: self-blame, other-blame, rumination, catastrophising, putting into perspective positive refocusing, positive reappraisal, acceptance, and planning. Each subscale is measured by four items and has possible scores ranging from 4 till 20. \cite{GarnefskiKraaij2007} have two criterion variables: depressive and anxiety symptoms as measured by the SCL-90. We will use the depression subscale which is measured by 16 items with possible scores between 16 and 80. 

We compare two lasso regression models, one standard model where the coefficients of the predictor variables are shrunken towards zero, and the reparametrization such that coefficients are shrunken towards equality, i.e. model \ref{eq:constlasso}. We used 10 fold cross-validation as imposed in the glmnet package \citep{glmnet}. 

\begin{figure}
\begin{center}
\includegraphics[width = 0.9\textwidth]{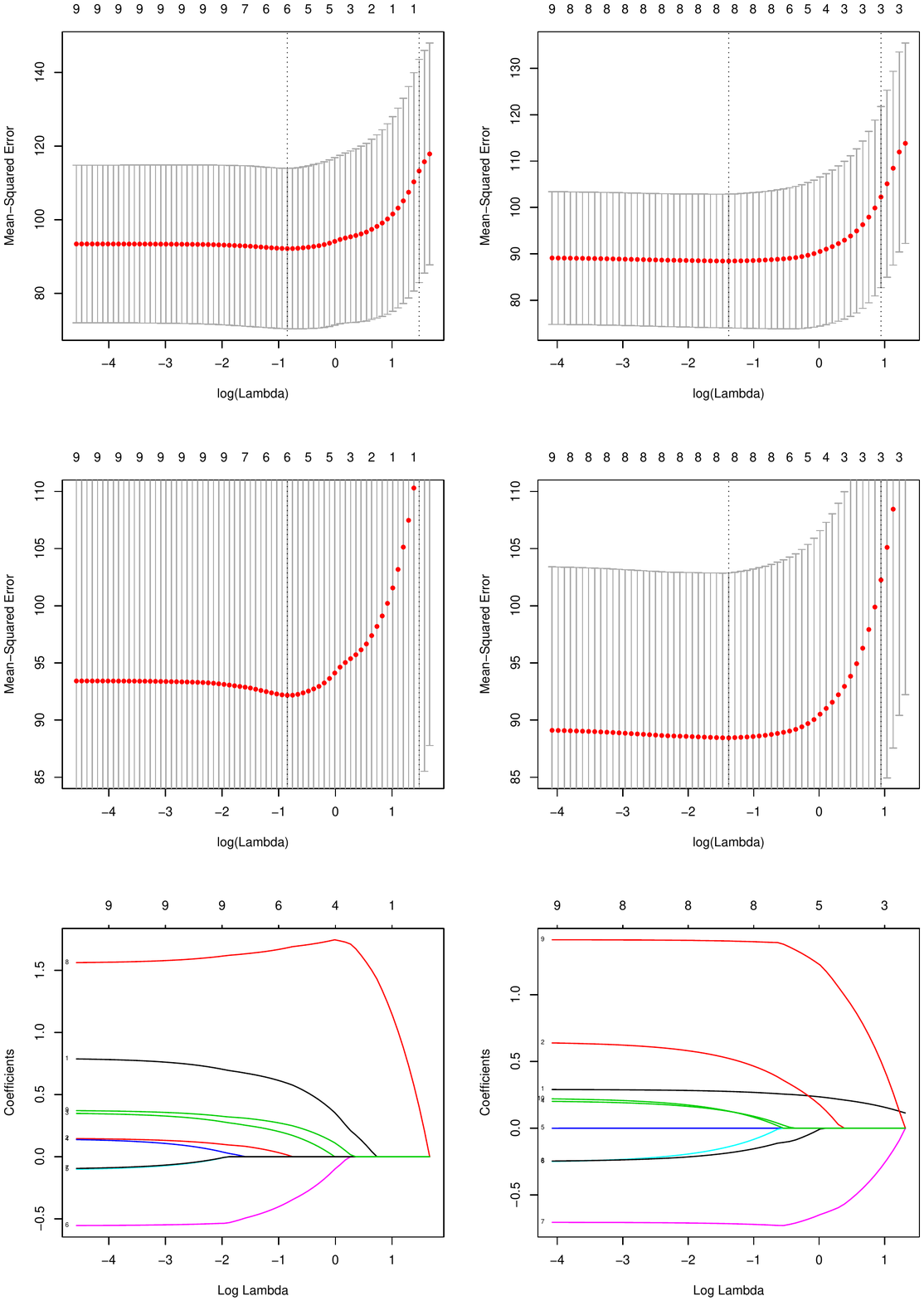}
\end{center}
\caption{Results of two lasso models on CERQ data. Left: standard lasso model with shrinkage to zero. Right: lasso with shrinkage towards equal regression coefficient. The upper plots give the 10 fold cross-validated mean squared errors. The middle plot provide the same information, but is zoomed such that the two models can be better compared. The vertical lines indicate the values of the penalty parameter with the lowest cross-validated error and with a cross-validated error within one standard error of the minimum. The lower plots provide plots of the regression weights given the value of the (log of the) penalty parameter. }
\label{fig:cerq}
\end{figure}

The results of fitting the model in the calibration set are shown in Figure \ref{fig:cerq} where on the left hand side the results are given for the standard lasso model and on the right hand side for the lasso as defined in equation \ref{eq:constlasso}. In the 10-fold cross-validation plots (upper and middle row) we see that the mean squared error for the newly proposed model is somewhat lower than for the standard lasso model. In the lower plots we see the regression coefficients as functions of the penalty parameter. 

In Figure \ref{fig:cerq} we see that the mean squared errors obtained by shrinkage towards equal coefficients is lower than these from the standard lasso (most clearly seen in the middle rows of the figure). The minimum mean squared error for the standard lasso model is 92.16, whereas that for the new parametrization is 88.45. Table \ref{tab:cerqcoef} shows the estimated regression coefficients in the standard regression parametrization and the parametrizatoin of Equation \ref{eq:constlasso}. 

\begin{table}
\centering
\begin{tabular}{r|rrr|rrr}
  \hline
 & OLS & \multicolumn{2}{c|}{lasso-zero}  &  OLS & \multicolumn{2}{c}{lasso-equal} \\ 
 &  & min & 1se &  & min & 1se \\ 
  \hline
Intercept & 9.32 & 10.77 & 22.27 &9.32 &  9.66 & 13.00 \\ 
  Sum & - & - & - & 0.29 & 0.28 & 0.16 \\ 
  Self-blame & 0.79 & 0.59 & 0 & 0.42 & 0.52 & 0 \\ 
  Acceptance & 0.15 & 0.01 & 0 & -0.22 & 0 & 0 \\ 
  Rumination & 0.35 & 0.18 & 0 & -0.02 & 0.13 & 0 \\ 
  Positive Refocusing & 0.15 & 0 & 0 & -0.23 & 0 & 0 \\ 
  Refocus on Planning & -0.11 & 0 & 0 & -0.48 & -0.14 & 0 \\ 
  Positive Reappraisal & -0.56 & -0.37 & 0 & -0.93 & -0.71 & -0.30 \\ 
  Putting into Perspective & -0.10 & 0 & 0 & -0.48 & -0.18 & 0 \\ 
  Catastrophising & 1.56 & 1.68 & 0.40 & 1.18 & 1.40 & 0.50 \\ 
  Blaming Others & 0.37 & 0.26 & 0 & - & 0.13 & 0 \\ 
   \hline
\end{tabular}
\caption{Estimated regression weights for the two models using no penalty (0; i.e., OLS coefficients), a penalty having minimum cross-validated error (min), and the one that falls within 1 standard error (1se). In the columns under Lasso-zero we used equation \ref{eq:regression} where the coefficients are shrunken to 0; In the columns under Lasso-equal we used equation \ref{eq:constlasso} where the regression coefficients are shrunken to be equal.}
\label{tab:cerqcoef}
\end{table}

\section{Conclusions and Discussion}

Machine learning methods become more and more popular for psychological research. The main aim of these models is to find prediction rules that have good predictive performance. Predictive performance is often measured using cross-validation techniques. Compared to linear models the prediction rules are based on statistical models that are either more or less flexible. More flexibility is obtained by basis expansions, whereas less flexibility is obtained by regularization. 

We showed that the field of psychometrics considered cross-validation and regularization already at the beginning of the 20th century as viable approaches to obtain generalizable results, i.e. prediction rules that would achieve better performance in practice. As such we can say that psychometrics is really at the origin of statistical learning theory. 

On the other hand, it is sad to see how little of this rich history is portrayed in our common wisdom. A standard psychometric book covers the topic of predictive validity, but (as far as we know) does not cover regularization, the bias variance trade-off, nor cross-validation. Hence, many psychologist keep on studying predictive validity in a suboptimal manner (using in-sample estimates) and provide overly optimistic estimates of predictive validity. We therefore suggest that this theory is re-incorporated in basic psychometric textbooks. 

Based on our review of the early psychometric literature we wondered about two issues. 
\begin{itemize}
\item In psychometric textbooks it is often written that with decreasing reliability of a test the predictive validity also goes down. Given that the unreliability of a test score find its way to the error in a regression model we concluded that more shrinkage is needed for unreliable tests. We tested this in a simulation study, and found that this is indeed correct. Furthermore, in this simulation study we found that there is a very weak relationship between reliability and predictive accuracy (see Figure \ref{fig:fig3b}) for small effect sizes. When the effect size is larger the relationship between reliability and predictive accuracy becomes stronger.

This finding has an important consequence for test assessment. The Dutch Committee for Test Evaluation (COTAN), for example, uses the criterion that the reliability of a test should be at least .90 in order to be used as a selection test. Such a criterion is based on the idea that predictive validity goes up with reliability. We showed, that this is not a linear relationship and that for weak effects sizes, as often found in psychology, measurement error hardly influences predictive accuracy. So, to assess psychological tests that are used for selection it is important to explicitly focus on prediction error or accuracy instead of using the surrogate of reliability. 
\item The second issue is that in early psychometrics regularization was often in terms of an equal or even unit regression weight for items in a test or tests in a test battery. Modern regularizarion techniques often shrink coefficients towards zero, not towards the mean coefficient. We rewrote the linear regression model in terms of a sum score and item deviation scores, i.e. the item score minus the person average over the items. Using this rewritten model we are able to shrink towards an equal regression weight for all items. Using an empirical example we showed that such a regularization might indeed be beneficial.   
\end{itemize}

We would like to notice two more things here. \cite{Mosier1951} discussed \emph{validity generalization}, the question whether a regression equation derived on a sample from one population generalizes to a second sample from another population. Such a second population may differ from the first population in terms of location (i.e. regression equation derived in Europe, validation sample from Asia) or time (regression equation derived in 2000, validation sample in 2019). Usual cross-validation techniques, like $K$-fold cross-validation nowadays often empowered, do not test for this type of generalization and therefore even cross-validated results obtained on a sample should always be taken with a grain of salt. 

\cite{Darlington68} already noted that the well known relationship between mean squared error and multiple $R$ does not hold out-of-sample. Making out-of-sample predictions the mean value is not calibrated and therefore the usual in-sample relationships between mean squared error, variance explained, and correlation are no longer true. We might obtain, for example, that for new observations with observed criterion values $(5,6,7,8,9)$ one prediction rule gives predictions $(1,2,3,4,5)$ and another one $(6,6.5,7,7.5,8)$. If we would use the correlation between predicted value and observed value as measure of predictive validity both sets of predictions would have predictive validity equal to 1. The second set of predictions is, however, much better than the first, which would be evident by using average squared difference. More information is given in \cite{Alexanderea2015}. \cite{Darlington68} wrote: "The correlation coefficient is more useful in ``fixed quota'' situations, and the mean square error is more useful in ``flexible quota'' situations". We would like to add that a test is often used for single person decisions and in such a case having a correlation measure is not helpful.

\section*{Acknowledgement}
We thank Nadia Garnefski and Vivian Kraaij for providing us with the empirical data. 
% \section*{Author Contributions}
%MdR and ED generated the idea of the study. MdR programmed the simulation studies in R. BCP requested the empirical data set and helped creating the Figures in R. MdR and HK wrote the two more technical boxes. MdR wrote the first draft of the manuscript. HK, MF, and ZB critically edited it. All authors approved the final submitted version of the manuscript. 
% \section*{Conflict of Interest}
% The authors declared that there were no conflicts of interest with respect to the authorship or the publication of this article. 
\section*{Open practices}
The R code for the experiment and for the analysis of the empirical data can be requested from the first author. 

%\bibliography{/users/rooijmjde/surfdrive/predictive-psychometrics/paper/predpsycho.bib}
%\bibliography{/users/mdr/surfdrive/predictive-psychometrics/paper/predpsycho.bib}

\begin{thebibliography}{}

\bibitem[Alexander et~al., 2015]{Alexanderea2015}
Alexander, D. L.~J., Tropsha, A., and Winkler, D.~A. (2015).
\newblock Beware of $r^2$: simple, unambiguous assessment of the prediction
  accuracy of qsar and qspr models.
\newblock {\em Journal of Chemical Information and Modeling}, 55:1316--1322.

\bibitem[Berk, 2008]{Berk2008}
Berk, R. (2008).
\newblock {\em Statistical learning from a regression perspective}.
\newblock Springer, New York.

\bibitem[Browne, 1975a]{Browne1975b}
Browne, M.~W. (1975a).
\newblock A comparison of single sample and cross-validation methods for
  estimating the mean squared error of prediction in multiple linear
  regression.
\newblock {\em British Journal of Mathematical and Statistical Psychology},
  28:112--1120.

\bibitem[Browne, 1975b]{Browne1975a}
Browne, M.~W. (1975b).
\newblock Predictive validity of a linear regression equation.
\newblock {\em British Journal of Mathematical and Statistical Psychology},
  28:79 -- 87.

\bibitem[Burt, 1950]{Burt1950}
Burt, C. (1950).
\newblock The influence of differential weighting.
\newblock {\em British Journal of Psychology, Statistical Section}, 3:105F --
  125.

\bibitem[Chapman et~al., 2016]{Chapmanetal2016}
Chapman, B.~P., Weiss, A., and Duberstein, P.~R. (2016).
\newblock Statistical learning theory for high dimensional prediction:
  Application to criterion-keyed scale development.
\newblock {\em Psychological Methods}, 21:603--620.

\bibitem[Claudy, 1978]{Claudy1978}
Claudy, J.~G. (1978).
\newblock Multiple regression and validity estimatoin in one sample.
\newblock {\em Applied Psychological Measurement}, 2:595--607.

\bibitem[Cureton, 1951]{Cureton1951}
Cureton, E.~E. (1951).
\newblock Approximate linear restraints and best predictor weights.
\newblock {\em Educational and Psychological Measurement}, 11:12--15.

\bibitem[Darlington, 1968]{Darlington68}
Darlington, R. (1968).
\newblock Multiple regression in psychological research and practice.
\newblock {\em Psychological Bulletin}, 69:161 -- 182.

\bibitem[Darlington, 1978]{Darlington78}
Darlington, R. (1978).
\newblock Reduced-variance regression.
\newblock {\em Psychological Bulletin}, 85:1238 -- 1255.

\bibitem[Freeman, 1926]{Freeman1926}
Freeman, F.~N. (1926).
\newblock {\em Mental Tests: Their History, Principles and Applications}.
\newblock Houghton Mifflin, Oxford, England.

\bibitem[Friedman et~al., 2010]{glmnet}
Friedman, J., Hastie, T., and Tibshirani, R. (2010).
\newblock Regularization paths for generalized linear models via coordinate
  descent.
\newblock {\em Journal of Statistical Software}, 33(1):1--22.

\bibitem[Garnefski and Kraaij, 2007]{GarnefskiKraaij2007}
Garnefski, N. and Kraaij, V. (2007).
\newblock The cognitive emotion regulation questionairre: Psychometric features
  and prospective relationships with depression and anxiety in adults.
\newblock {\em European Journal of Psychological Assessment}, 23:141 -- 149.

\bibitem[Guilford, 1936]{Guilford1936}
Guilford, J.~P. (1936).
\newblock {\em Psychometric methods}.
\newblock McGraw-Hill, New York.

\bibitem[Hastie et~al., 2009]{HTF2009}
Hastie, T., Tibshirani, R., and Friedman, J. (2009).
\newblock {\em The elements of statistical learning, 2nd edition}.
\newblock Springer, New York.

\bibitem[Hastie et~al., 2015]{Sparsity2015}
Hastie, T., Tibshirani, R., and Wainwright, M. (2015).
\newblock {\em Statistical Learning with Sparsity: The Lasso and
  Generalizations}.
\newblock Chapman and Hall/CRC, Boca Raton, FL.

\bibitem[Herzberg, 1969]{Herzberg1969}
Herzberg, P.~A. (1969).
\newblock The parameters of cross-validation.
\newblock {\em Psychometrika}, 34:Monograph.

\bibitem[Hoerl and Kennard, 1970]{HoerlKennard1970}
Hoerl, A.~E. and Kennard, R.~W. (1970).
\newblock Ridge regression: Biased estimation for nonorthogonal problems.
\newblock {\em Technometrics}, 12:55 -- 67.

\bibitem[Larson, 1931]{Larson1931}
Larson, S.~C. (1931).
\newblock The shrinkage of the coefficient of multiple correlation.
\newblock {\em Journal of Educational Psychology}, 22:45--55.

\bibitem[Lawshe and Schucker, 1959]{LawsheSchucker1959}
Lawshe, C. and Schucker, R. (1959).
\newblock The relative efficiency of four test weighting methods in multiple
  prediction.
\newblock {\em Ediucational and Psychological Measurement}, 14:103 -- 114.

\bibitem[McNeish, 2015]{McNeish2015}
McNeish, D.~M. (2015).
\newblock Using lasso for predictor selection and to assuage overfitting: A
  method long overlooked in behavioural sciences.
\newblock {\em Multivariate Behavioural Research}, 50:471 -- 484.

\bibitem[Mosier, 1951]{Mosier1951}
Mosier, M.~W. (1951).
\newblock Problems and design of cross-validation.
\newblock {\em Educational and Psychological Measurement}, 11:5--11.

\bibitem[Pruzek and Frederick, 1978]{Pruzek78}
Pruzek, R. and Frederick, B. (1978).
\newblock Weighting predictors in linear models: Alternatives to least squares
  and limitations of equal weights.
\newblock {\em Psychological Bulletin}, 85:254 -- 266.

\bibitem[Putka et~al., 2018]{Putkaea2018}
Putka, D.~J., Beatty, A.~S., and Reeder, M.~C. (2018).
\newblock Modern prediction methods: new perspectives on common problems.
\newblock {\em Organizational Research Methods}, 21:689 -- 732.

\bibitem[Rozeboom, 1978]{Rozeboom1978}
Rozeboom, W.~W. (1978).
\newblock Estimation of cross-validated multiple correlation: A clarification.
\newblock {\em Psychological bulletin}, 85:1348--1351.

\bibitem[Schmidt, 1971]{Schmidt1971}
Schmidt, F. (1971).
\newblock The relative efficiency of regression and simple unit predictor
  weights in applied differential psychology.
\newblock {\em Educational and Psychological Measurement}, 31(3):699--714.

\bibitem[Tibshirani, 1996]{Tibshirani1996}
Tibshirani, R. (1996).
\newblock Regression shrinkage and selection via the lasso.
\newblock {\em Journal of the Royal Statistical Society, B}, 58:267 -- 288.

\bibitem[Tonidandel et~al., 2016]{Tonidandelea2016}
Tonidandel, S., King, E.~B., and Cortina, J.~M. (2016).
\newblock {\em Big data at work: The data science revolution and organizational
  psychology}.
\newblock Routledge, New York, NY.

\bibitem[Wainer, 1976]{Wainer76}
Wainer, H. (1976).
\newblock Estimating coefficients in linear models: It don't make no nevermind.
\newblock {\em Psychological Bulletin}, 83:213 -- 217.

\bibitem[Wherry, 1951]{Wherry1951}
Wherry, R.~J. (1951).
\newblock Iv. comparison of cross-validation with statistical inference of
  betas and multiple $r$ from a single sample.
\newblock {\em Educational and Psychological Measurement}, 11:23--28.

\bibitem[Yarkoni and Westfall, 2017]{YarkoniWestfall2017}
Yarkoni, T. and Westfall, J. (2017).
\newblock Choosing prediction over explanation in psychology: Lessons from
  machine learning.
\newblock {\em Perspectives on Psychological Science}, 12:1100 -- 1122.

\bibitem[Zou and Hastie, 2005]{ZouHastie2005}
Zou, H. and Hastie, T. (2005).
\newblock Regularization and variable selection via the elastic-net.
\newblock {\em Journal of the Royal Statistical Society, B}, 67:301 -- 320.

\end{thebibliography}
%\bibliographystyle{apalike}

\end{document}